\documentclass{article}

\usepackage[preprint]{neurips_2024}

\usepackage[utf8]{inputenc} 
\usepackage[T1]{fontenc}    
\usepackage{hyperref}       
\usepackage{url}            
\usepackage{booktabs}       
\usepackage{amsfonts}       
\usepackage{nicefrac}       
\usepackage{microtype}      
\usepackage{xcolor}         
\usepackage{amsmath}
\usepackage{graphicx}

\usepackage{amsmath}

\title{Physics-Informed Neural Networks for Solving the Two-Dimensional Shallow Water Equations with Terrain Topography and Rainfall Source Terms}

\author{%
  Yongfu Tian\(^{1,2}\) \quad Shan Ding\(^{1,2}\) \quad Guofeng Su\(^{1,2} \) \quad Lida Huang\(^{1,2}\) \quad Jianguo Chen\(^{1,2}\)  \\
  $^1$School of Safety Science,
  Tsinghua University\\
  $^2$Key Laboratory of Investigation and Evaluation on Natural Hazards and Industrial Accidents,\\ Ministry of Emergency Management, Tsinghua University\\
  \texttt{tianyf22@mails.tsinghua.edu.cn} \\
  \texttt{chenjianguo@mail.tsinghua.edu.cn} \\
}

\begin{document}

\maketitle

\begin{abstract}
  Solving the two-dimensional shallow water equations is a fundamental problem in flood simulation technology. In recent years, physics-informed neural networks (PINNs) have emerged as a novel methodology for addressing this problem. Given their advantages in parallel computing, the potential for data assimilation and parameter calibration, and the rapid advancement of artificial intelligence, it is crucial to investigate both the capabilities and limitations of PINNs. While current research has demonstrated the significant potential of PINNs, many aspects of this new approach remain to be explored. In this study, we employ PINNs enhanced by dimensional transformation and N-LAAF techniques to validate their effectiveness in solving two-dimensional free surface flow with rainfall on terrain topography. The shallow water equations primarily exist in two forms: the variables form and the conservative form. Through theoretical analysis and experimental validation, we demonstrate that a hybrid variable-conservation form offers superior performance. Additionally, we find that incorporating the energy conservation law, specifically the entropy condition, does not yield substantial improvements and may even lead to training failure. Furthermore, we have developed an open-source module on the PINNacle platform for solving shallow water equations using PINNs, which includes over ten case studies and various equation forms, to promote research and application in this field.
\end{abstract}

\section{Introduction}
With the intensification of global climate change, the frequency and severity of flood events are increasing, leading to escalating losses from floods \citep{RN376}. Consequently, the prevention and mitigation of flood disasters have become increasingly critical. One of the key challenges in this context is the accurate simulation and prediction of floods \citep{RN397}. In addition to numerical-driven models, solutions over the past few decades have primarily relied on Shallow Water Equations (SWE) or their simplified forms, such as the kinematic wave equation and the diffusion wave equation.

Except for special cases such as the Riemann problem \citep{RN390}  and one-dimensional flood wave propagation problems in planar geometries \citep{RN391}, obtaining analytical solutions to the shallow water equations is challenging. Therefore, numerical methods are typically employed. Over the past few decades, a wide range of numerical methods for solving SWE have been developed, which can be primarily categorized into three classes: finite difference methods \citep{RN392}, finite volume methods \citep{RN393,RN416}, and finite element methods \citep{RN394}. Benefiting from advancements in CPU computing power and, to some extent, GPU computing capabilities, these methods have been extensively applied to real-world flood prediction problems, achieving excellent results.

In recent years, the rapid advancement of artificial neural networks has led to the emergence of physics-informed neural networks (PINNs) as an alternative approach to solving partial differential equations (PDEs), distinct from traditional numerical methods \citep{RN399}. PINNs approximate the solutions of PDEs using neural network functions, where the inputs are the independent variables of the PDEs and the outputs are the dependent variables. To integrate physical knowledge, PINNs embed the governing PDEs and boundary conditions into the loss function of the neural network. By leveraging backpropagation \citep{RN400} and gradient-based optimization algorithms, the loss function is minimized, thereby making the neural network's output increasingly accurate in approximating the true solution of the PDEs. This process constitutes an unsupervised learning algorithm. Additionally, observational data can be incorporated into the loss function, enabling PINNs to simultaneously approximate both the PDE solutions and the observational data, thus achieving data assimilation \citep{RN401} —a task that remains challenging for conventional numerical models.

As summarized by Dazzi \citep{RN378}, the key advantages of the PINN method include its mesh-free nature, ease of incorporating observational data, simplicity in parameter calibration, and flexibility in calibrating boundary conditions (such as upstream inflows). However, the application of PINNs to solve SWE is still in its exploratory phase, and several technical challenges remain. These challenges include the well-balance of the different components of the loss function \citep{RN375}, difficulties in training, and long training times. Moreover, the overall accuracy of PINNs is currently lower than that of numerical methods. Fortunately, the accuracy requirements for simulating flood events are not extremely stringent. Despite these limitations, PINNs offer a promising new methodology that warrants further exploration of their capabilities and limitations, providing additional tools for solving shallow water equations. With ongoing advancements in GPU computing power, neural network architectures, and optimization algorithms, we anticipate that PINNs will play a significant role in future flood prediction.

Currently, researchers have conducted initial explorations into solving SWE using PINNs, primarily focusing on relatively simple one-dimensional (1D) cases, while studies on more complex two-dimensional (2D) scenarios remain limited. Cedillo et al. \citep{RN407} applied PINNs to solve a one-dimensional steady-state river flow problem. To achieve convergence, they performed dimensional transformation on the energy differential equation, which is a steady-state simplification of SWE. Their results demonstrated that PINN can effectively simulate different flow states in complex riverbeds. Feng et al. \citep{RN409} embedded PINNs into the sub-grid of numerical schemes and incorporated various observational data into PINNs to enhance the accuracy of river flow problems. Yin et al. \citep{RN412} used PINNs to solve the one-dimensional unsteady SWE and applied them to Cypress Creek downstream of Houston, obtaining results consistent with numerical solutions. Dazzi \citep{RN378} proposed a novel augmentation method that integrates topographical information directly into the neural network. The experimental results demonstrate that this approach can not only accurately predict water depth and flow velocity but also precisely characterize the terrain. Overall, the capability of the PINN method to solve the 1D SWE has been well validated. It demonstrates proficiency in handling complex terrains and various flow states in one dimension, although many challenges still need further investigation.

To advance the engineering application of PINNs in addressing rainstorm flood problems, we need to focus our research on the 2D SWE. Preliminary studies have already been conducted for the 2D cases. Leiteritz et al. \citep{RN413} explored the effectiveness of using different optimizers in PINNs to solve the radial dam break problem without topography, demonstrating that PINNs can achieve accuracy comparable to traditional numerical methods. Qi et al. \citep{RN414} employed both fully connected neural networks and convolutional neural networks to solve 2D SWE and applied these models to a real river basin. The simulation results were found to be comparable to those obtained using the finite volume method. Bihlo and Popovych \citep{RN411} focused on the SWE on a spherical surface in meteorology, proposing to train a series of neural networks over the integration interval to address the challenges of long-term simulations. 
These advancements have highlighted the significant potential of PINNs in solving 2D SWE. However, as a novel methodology, PINNs still present several important issues that warrant further exploration. This study focuses on the following key questions:

1. Rainstorm and flood problems introduce the rainfall source terms and complex terrain topologies into SWE. Can PINNs effectively solve SWE with rainfall source terms and complex terrain topologies?

2. There are various forms of SWE. Which form is most suitable for embedding into PINNs?

3. SWE encompass the conservation laws of mass and momentum. Would incorporating the conservation law of energy improve the model's performance?

Regarding the first question, we initially set up three static water cases with terrain topography. The results demonstrated that PINNs can effectively address the "well-balance" problem of static water, which is particularly challenging for numerical methods. Subsequently, we establish four static water cases with rainfall and terrain topography, showing that PINNs can simulate rainfall processes with high accuracy. Finally, we conduct four dynamic cases: the 2D dam break problem, the 2D circular dam break problem, the tidal problem, and the tidal problem with rainfall. Comparisons with exact solutions or finite volume numerical solutions reveal that PINNs can effectively handle shock discontinuities and flow problems with rainfall on complex terrains.

For the second and third questions, our motivation stems from the numerical methods used for solving SWE. SWE can be expressed in conservative form and primitive variables form, with the former typically solved using finite volume methods and the latter using finite difference methods. It is crucial to explore how these two forms perform within the PINN framework. Our theoretical and experimental analyses indicate that a hybrid variable-conservation form offers superior performance. 
In numerical methods, the energy conservation law serves as an entropy condition to manage shock discontinuities and enhance the performance of numerical schemes. While PINNs gradually approximate the solution of SWE through optimization techniques, they do not inherently satisfy the energy conservation. To ensure the approximate solution meets the energy conservation condition, we introduce an entropy condition. Unexpectedly, our case studies show that incorporating the energy condition does not provide significant advantages and may even lead to training instability.

It is well known that training PINNs is rather challenging. Wang et al. \citep{RN376,RN375} identified an imbalance among the loss terms of PINNs and stiffness issues in the gradient flows of each term, leading to optimization difficulties. Additionally, SWE exhibit substantial discrepancies in the magnitudes of various physical quantities, such as bottom topography and water depth, which further contribute to the inherent imbalance of the equations. Krishnapriyan et al. \citep{RN374} highlighted that the training performance of PINNs is highly sensitive to the coefficients of partial differential equations. To effectively train SWE with source terms and topography, this study employs dimensionality transformation technique and the N-LAAF technique \citep{RN377}. The results demonstrate that these two techniques can significantly enhance the training of PINNs.

Although PINNs for SWE have attracted increasing attention, currently only a few research codes and data are currently open source. Some open-source research can be referred to the works of Dazzi \citep{RN378} and Feng et al \citep{RN409}. In this study, a general module for solving SWE is constructed on the PINNacle platform \citep{RN418}. This module can calculate SWE with various topographies and rainfall source terms. Additionally, to verify the results of PINNs, two numerical solvers are paired with this platform, namely the HLL scheme \citep{RN416} and the entropy stable scheme \citep{RN384}, both of which can be accelerated by GPU. 

The remainder of this study is organized as follows: Section 2 introduces the PINN method and its application to SWE. It also provides a theoretical analysis of the advantages and disadvantages of various equation forms and discusses the dimensional transformation of SWE. Section 3 details the 11 case studies and the parameter settings for the models. Section 4 presents the results of these cases and provides a thorough discussion.

\section{Method}
\subsection{PINN}
Consider a set of partial differential equations defined on a bounded closed domain \(\Omega \subseteq R^n \)
\begin{align}
\mathcal{F}(u(x);x)=0,   x\in \Omega \\
\mathcal{B}(u(x);x)=0,   x \in \partial \Omega
\end{align}
where \(\mathcal{F}\) is the differential operator of the PDEs, \(\mathcal{B}\) denotes the boundary conditions, and \(x\) is the vector of independent variables (such as spatiotemporal coordinates). Assuming that the PDEs are well-posed, the true solution is \(u(x)\). And PINNs use a neural network with parameters \( \theta\), denoted as \(u_{\theta}  (x) \), to approximate \(u (x) \). To approach the true solution, PINNs embed the PDEs into their loss function, that is
\begin{equation}
\mathcal{L}(\theta)=\frac{\omega_f}{N_f} \sum_{i=1}^{N_f}\left\|\mathcal{F}\left(u_{\theta}\left(x_i\right)\right)\right\|^2+\frac{\omega_b}{N_b} \sum_{j=1}^{N_b}\left\|\mathcal{B}\left(u_{\theta}\left(x_j\right)\right)\right\|^2+\frac{\omega_d}{N_d} \sum_{k=1}^{N_d}\left\|u_{\theta}\left(x_k\right)-u\left(x_k\right)\right\|^2
\end{equation}
Here, \( \mathcal{L}(\theta) \) denotes the loss function of PINNs. \(N_f\) and \(N_b\) represent the numbers of training points randomly sampled from the domain \(\Omega\) and its boundary \(\partial \Omega\), respectively. \(N_d\) is the number of observational data points. If no observational data are available, this term is omitted. \(\omega_f\), \(\omega_b\) and \(\omega_d\) are the weights assigned to each loss term. During the process of minimizing the loss function, PINNs gradually converge to the true solution. To prevent over-fitting or accelerate training, regularization terms, such as \(\| \theta \|^2 \), may be added to the loss function.

Here, we briefly explain the fundamental principles of PINNs and analyse their errors. For simplicity and ease of understanding, the mathematical rigor is weakened. For convenience, we assume that \(\mathcal{B}\) acts as 0 on non-boundary points, i.e. \(\mathcal{B} (u;x)=0\) for \(x\notin \partial \Omega \). Assume \( u\in C^\infty \), where \(C^\infty\) is the vector space composed of smooth functions. Consider \( v\in C^\infty \) such that for \(\forall x \in \Omega \), \(\mathcal{F} (v;x)= \mathcal{F} (u;x)=0 \) and \(\mathcal{B} (v;x) = \mathcal{B} (u;x)=0 \). By the well-posedness of PDEs, it follows that \(v=u\).

We randomly sample \(N\) points within \(\Omega\) with a sampling density \(\rho\), where \(\rho\) is defined as the number of sampling points per unit volume. We select \(v_N \in C^\infty \) such that for each point \(x_i (1\leq i\leq N) \), \(\mathcal{F} (v_i;x_i ) = \mathcal{F} (u_i;x_i )=0 \) and \(\mathcal{B} (v_i,x_i) = \mathcal{B} (u_i,x_i )=0\). Consequently, as \(N \to \infty \) or \(\rho \to \infty \), \(v_N \to u\). 

The universal approximation theorem of neural networks states that a sufficiently wide \(u_{\theta} \) can approximate any smooth function. Assuming that \(u_{\theta} \) is smooth with respect to the input variables, the function space formed by \(u_{\theta} \) is \(C_{\theta} \subseteq C^\infty  \). If for each point \(x_i (1 \leq i \leq N)\), \(\mathcal{F}(u_{\theta};x_i) \to \mathcal{F}(u_i;x_i) = 0 \)  and \(\mathcal{B}(u_{\theta};x_i) \to \mathcal{B} (u_i;x_i )=0\), then as \(N \to \infty\) or \(\rho \to \infty \), we have \(u_{\theta} \to u \).

However, in practice, the number of sampling points and the number of parameters in \(u_{\theta} \) are always finite. Consequently, approximating \(u\) with \(u_{\theta} \) inevitably introduces errors. From the preceding analysis, it is evident that the approximation error is influenced by both the sampling density and the network's expressive power. Given a sampling density \(\rho\) and a neural network parameter space \(\theta\), the minimum error of PINNs is defined as \(\varepsilon_{\theta,\rho} = inf \|u-C_{\theta} \| \). As \(\rho \to \infty \) and the number of parameters in \(\theta\) tends to infinity, \(\varepsilon_{\theta,\rho} \to 0\). Additionally, the optimization problem of PINNs is non-convex, and the gradient descent algorithm may not converge to the global optimum. Consequently, there is an optimization error \(\varepsilon_{opt}= \| u_{\theta}-u_{{\theta},min} \|\), where \(u_{{\theta},min} =argmin\|u-C_{\theta} \| \). In summary, the total approximation error \(\varepsilon\) of \(u_{\theta}\) satisfies
\begin{equation}
  \varepsilon \leq \varepsilon_{\theta,\rho} +\varepsilon_{opt}  
\end{equation}

\subsection{PINN for SWE}
It is well known that the motion of water can be fully described by the incompressible Navier-Stokes (NS) equations. Under the assumptions of inviscid flow, hydrostatic pressure, uniform horizontal velocity, and bottom boundary conditions and free surface conditions, the NS equations can be simplified and derived into the shallow water equations \citep{RN389}. Mathematically, the SWE belong to the hyperbolic conservation law equations. After adding the rainfall source term and ground friction term, the conservation form of the two-dimensional shallow water equations is
\begin{equation}
    U_t+F(U)_x+G(U)_y=S_z+S_f  \label{swe}
\end{equation}
where
\begin{align}
U&=(h,hu,hv)^T \\
F(U)&=(hu,hu^2+\frac{1}{2} gh^2,huv)^T \\
G(U)&=(hv,huv,hv^2+\frac{1}{2} gh^2 )^T \\
S_z&=(p,-ghz_x,-ghz_y )^T \\
S_f&=\left(0,-Cu\sqrt{u^2+v^2},-Cu\sqrt{u^2+v^2} \right)^T
\end{align}
Here, \(h\) denotes the water depth, \(u\) and \(v\) represent the velocity components in the \(x\) and \(y\) directions, respectively. \(U\) is the vector of conserved quantities, \(F(U)\) and \(G(U)\) are the flux vectors of \(U\) in the \(x\) and \(y\) directions, respectively. \(S_z\) represents the source term due to topography and rainfall, while \(S_f\) denotes the friction source term. \(z\) is the topography function, with \(z_x\) and \(z_y\) being its partial derivatives with respect to \(x\) and \(y\). \(C=gn^2/h^{1/3}\), where \(n\) is the Manning’s coefficient, and \(p\) is the rainfall intensity. The friction term \(S_f\) is not considered in this study.

Based on the introduction of PINNs in Section 2.1, we can incorporate SWE directly into the loss function of the neural network to formulate a PINN. Let the neural network be denoted as \(U_{\theta}\), with inputs \((x,y,t)\) and outputs \(( h_{\theta},u_{\theta},v_{\theta} )\). The loss function \(\mathcal{L}\) for \(U_{\theta}\) is defined as
\begin{equation}
    \mathcal{L}=\frac{\omega_e}{N} \mathcal{L}_E+\frac{\omega_I}{N_I} \mathcal{L}_I +\frac{\omega_B}{N_B} \mathcal{L}_B
\end{equation}
where each term is specified as follows
\begin{align}
\mathcal{L}_E &=\sum_{i=1}^N {\| (U_{\theta,i} )_t+F(U_{\theta,i} )_x+G(U_{\theta,i} )_y-S_{z,i} \|_2^2 }, \quad (x_i,y_i,t_i ) \in \Omega\\
\mathcal{L}_I &=\sum_{j=1}^{N_I} { \|U_{\theta }(x_j,y_j,0) -U(x_j,y_j,0) \|_2^2}, \quad (x_j,y_j,0) \in \partial \Omega\\
\mathcal{L}_B &=\sum_{j=1}^{N_I}  \|U_{\theta }(x_j,y_j,t_j) -U(x_j,y_j,t_j) \|_2^2, \quad (x_j,y_j,t_j ) \in \partial \Omega
\end{align}
Among these, \(\omega_e\), \(\omega_I\), and \(\omega_B\) are weighting parameters, while \(N\), \(N_I\), and \(N_B\) represent the numbers of sample points in the interior domain, initial conditions, and boundary conditions, respectively. Note that \(N_B\) excludes the sample points from the initial conditions.

After constructing the standard PINN structure, we must address several specific challenges encountered when applying PINNs to solve SWE. In flow phenomena, the water depth must remain non-negative and satisfy dry-wet conditions (i.e., if the water depth is zero, then the velocity must also be zero). The PINN method using fitting and optimization approaches cannot inherently ensure these conditions and may even produce negative water depths, leading to optimization difficulties. To guarantee non-negative water depth and dry-wet conditions, we adopt the approach proposed by Dazzi \citep{RN378} and introduce three additional loss functions
\begin{align}
\mathcal{L}_{pos} &=\sum_{i=1}^{N}  \|h_{\theta }(x_i,y_i,t_i) - h_{\theta,i }\cdot Heaviside(h_{\theta,i }  ) \|_2^2, \quad (x_i,y_i,t_i ) \in  \Omega \\
\mathcal{L}_{dry,u} &=\sum_{i=1}^{N}  \|u_{\theta }(x_i,y_i,t_i) - u_{\theta,i }\cdot Heaviside(h_{\theta,i }  ) \|_2^2, \quad (x_i,y_i,t_i ) \in  \Omega    \\
\mathcal{L}_{dry,v} &=\sum_{i=1}^{N}  \|v_{\theta }(x_i,y_i,t_i) - v_{\theta,i }\cdot Heaviside(h_{\theta,i }  ) \|_2^2, \quad (x_i,y_i,t_i ) \in  \Omega
\end{align}
where the function \(y=Heaviside(x) \) is defined as: if \(x<0\), then \(y=0\), and if \(x\geq 0\), then \(y=1\).

PINN solves SWE through an iterative optimization process, and the approximate solutions obtained may violate the law of energy conservation. To further regularize these approximate solutions to satisfy the energy conservation, we introduce the entropy condition. Ignoring the source term on the right-hand side of Equation (\ref{swe}), the shallow water equations fall into the category of nonlinear hyperbolic conservation laws. It is well known that, regardless of whether the initial conditions are smooth or not, the solution \(U\) of SWE can develop shock discontinuities within a finite time. At these discontinuities, \(U\) is not differentiable and no longer satisfies the equations, although it still satisfies them in the non-discontinuity regions. To address this issue, Lax generalized the concept of solutions for differential equations and introduced the concept of weak solutions \citep{RN379}. While weak solutions allow for discontinuous solutions, they are not unique, contradicting the requirement for a unique physical solution. To select the physically meaningful unique solution from among the weak solutions, the entropy condition was introduced. The entropy condition ensures energy conservation in smooth regions but allows for energy dissipation in shock discontinuity regions.

When the ground friction term \(S_f\) and the rainfall source term \(p\) are not considered, the entropy conservation condition is
\begin{equation}
    E(U)_t+(H(U)+J_1 )_x+(K(U)+J_2 )_y=0
\end{equation}
Considering the possible energy dissipation of shock waves, the entropy stability condition is also considered, i.e.
\begin{equation}
    E(U)_t+(H(U)+J_1 )_x+(K(U)+J_2 )_y \leq 0
\end{equation}
where
\begin{align}
    E(U)&=\frac{1}{2} (hu^2+hv^2+gh^2 )+ghz \\
    H(U)&=\frac{1}{2} (hu^3+huv^2 )+gh^2 u \\
    K(U)&=\frac{1}{2} (hu^2 v+hv^3 )+gh^2 v \\
    J&=ghz (u,v)^T    
\end{align}

Ignoring the kinetic energy of rainfall and considering that rainfall only serves to increase water depth, the entropy stability condition with the rainfall source term is
\begin{equation}
    E(U)_t+(H(U)+J_1 )_x+(K(U)+J_2 )_y-g(h+b)p \leq 0
\end{equation}
Let the left-hand side term of the inequality be denoted as \(ER(U,p)\). Then, the loss term \(\mathcal{L}_{En}\) corresponding to the entropy stability condition is
\begin{equation}
    \mathcal{L}_{En}=\sum_{i=1}^N \| {ReLU \left(ER \left(U_{\boldsymbol{\theta,i}},p_i \right) \right) \|_2^2 }
\end{equation}
The ReLU function penalizes positive values of the entropy condition. In addition, the loss term for the entropy conservation condition does not require the ReLU function.

Based on the above analysis, after incorporating the entropy stability condition, the non-negativity condition of water depth, and the dry-wet condition, the complete loss function for \(U_{\theta}\) is given by
\begin{equation}
    \mathcal{L}_{\theta}= \frac{\omega_e}{N} \mathcal{L}_E+\frac{\omega_I}{N_I} \mathcal{L}_I +\frac{\omega_B}{N_B} \mathcal{L}_B  + \frac{\omega_{En}}{N} \mathcal{L}_{En} + \frac{\omega_{p}}{N} \mathcal{L}_{pos} +\frac{\omega_{d}}{N} \left(\mathcal{L}_{dry,u}+ \mathcal{L}_{dry,v} \right) \label{loss}
\end{equation}
where  \(\omega_e\), \(\omega_I\), \(\omega_B\), \(\omega_{En}\), \(\omega_p\) and \(\omega_d\) are the corresponding weight parameters.

\subsection{The selection of equation forms} \label{sec2_3}
SWE can be expressed in multiple forms, including the conservative form, primitive variables form, and the hybrid variable-conservation form. Different equation forms lead to variations in the loss term \(\mathcal{L}_E\) within the loss function (\ref{loss}) for PINNs. To determine which form is most suitable for PINNs, we will analyse the advantages and disadvantages of each form. For simplicity and without loss of generality, the following discussion excludes the rainfall terms.

The conservation form is presented in Equations (\ref{swe}). If the conserved quantities \((h,hu,hv)\) are used as outputs for PINNs, it becomes necessary to handle derivatives involving zero division, such as \((hu^2 )_x=((hu)^2/h)_x\) , to prevent the back-propagation algorithm from failing. It is important to note that this issue is unrelated to whether the water depth is always greater than zero, as the initialized parameters and those during the training process may result in \(h_{\theta} (x,y,t)=0\). Our preliminary experiments indicate that this approach leads to relatively unstable training. Variants of the conservation form, given by 
\begin{equation}
    U_t+A(U) U_x+B(U) U_y=S_z
\end{equation}
where \(A(U)=\nabla_U F(U)\) and \(B(U)=\nabla _U G(U)\), also require handling derivatives with zero division, such as \(u_{\theta}=(hu/h)_{\theta}\), which similarly results in instability.

The shallow water equations in primitive variables form are given by
\begin{align}
    h_t+h_x u+hu_x+h_y v+hv_y=0\\
    u_t+uu_x+vu_y+gh_x+gz_x=0\\
    v_t+uv_x+vv_y+gh_y+gz_y=0
\end{align}
Let \(r_1\), \(r_2\), and \(r_3\) denote the residuals of these three equations, respectively, such as \(r_1=h_t+h_x u+hu_x+h_y v+hv_y\). The residual vector is defined as \(r=(r_1,r_2,r_3 )^T\). 

The variable-conservation form of SWE is derived by fully expanding the conservation form of SWE (Equations \ref{swe}) into equations with \((h,u,v)\) as dependent variables. For example, \((hu)_x\) is expanded as \(h_x u+h u_x\). Let the residual vector of the variable-conservation form be \(R=(R_1,R_2,R_3 )^T\). The relationship between the residual vector \(R\) and the residual vector \(r\) of the primitive variables form is given by
\begin{equation}
    R=Ar
\end{equation}
where
\begin{equation}
A= \begin{bmatrix}
    1  & 0  & 0 \\
    u  & h  & 0 \\
    v  & 0  & h
\end{bmatrix}
\end{equation}

The 2-norm of the vector is taken. For any \(x \in \Omega \), the loss term \(\mathcal{L}_E \) of the primitive variables form and its gradient with respect to the parameter \(\theta\) are
\begin{align}
\| \mathcal{F}( u_{\theta }(x_i )) \| ^2 &= r^T r \\
\frac{1}{2} \nabla_{\theta} \| \mathcal{F}( u_{\theta }(x_i )) \| ^2 &= \left(\nabla_{\theta} r\right)^T  r
\end{align}
The loss term \(\mathcal{L}_E \) of the variable-conservation form is
\begin{equation}
    \| \mathcal{F}( u_{\theta }(x_i )) \| ^2 = \|R \|^2=\| Ar \|^2=r^T A^T Ar
\end{equation}
And its gradient with respect to the parameter \(\theta\) is
\begin{equation}
    \frac{1}{2} \nabla_{\theta} \| \mathcal{F}( u_{\theta }(x_i )) \| ^2 = \frac{1}{2} \nabla_{\theta} \|Ar\|^2 
    =(\nabla_{\theta} r)^T A^T Ar+ r^T (\nabla_{\theta} A)^T Ar  \label{grad}
\end{equation}
After training to a certain stage, \(r^T r\) will become smaller, making \(r^T (\nabla_{\theta} A)^T Ar\)  a relatively higher-order small term. Therefore, we first analyze the main term \((\nabla_{\theta} r)^T A^T Ar\). The matrix \(A^T A\) is symmetric and has non-negative real eigenvalues. Let its orthogonal decomposition be \(A^T A = Q \Lambda Q^T\), where \(\Lambda\) is a diagonal matrix composed of eigenvalues, and \(Q\) is an orthogonal matrix composed of right eigenvectors. Specifically,
\begin{equation}
    \Lambda=diag \left( h^2,0.5 \left(1+h^2+k^2 \pm
 \sqrt{\left((h+1)^2+k^2  \right) \left((h-1)^2+k^2 \right) }) \right) \right)\\
\end{equation}
where \(k^2=u^2+v^2\), and \(\pm\) indicates two terms. Let the three eigenvalues be \(\lambda_1\),\(\lambda_+\), and \(\lambda_-\), respectively. The corresponding orthogonal matrix \(Q\) composed of the eigenvectors is
\begin{equation}
    Q=\begin{bmatrix}
    0  & \frac{\lambda_+ - h^2}{C_+} & \frac{\lambda_- - h^2}{C_-} \\
    \frac{v}{k} &  \frac{hu}{C_+}  &\frac{hu}{C_-}\\
    -\frac{u}{k}  & \frac{hv}{C_+}  & \frac{hv}{C_-}  
    \end{bmatrix}
\end{equation}
where \(C_+\) and \(C_-\) are normalization factors defined as \(C_+=\sqrt{(\lambda_+-h^2 )^2+h^2 k^2 }\), \(C_-=\sqrt{(\lambda_--h^2 )^2+h^2 k^2 }\).

Furthermore, we have 
\begin{equation}
    (\nabla_{\theta}  r)^T A^T Ar=(Q^T \nabla_{\theta}  r)^T \Lambda (Q^T r)
\end{equation}
By comparing this with the gradient term \((\nabla_{\theta} r)^T r\) in the primitive variables form, it can be observed that the variable-conservation form first orthogonally transforms the gradient \((\nabla_{\theta} r)^T\) and the residual \(r\) into the characteristic field. Subsequently, it automatically weights these transformed quantities using the eigenvalues \(\Lambda\) in the characteristic field.

From an optimization perspective, we further examine the differences between the two formulations. Without loss of generality, consider any single point \(x \in \Omega\). Let \(\theta^*\) be the minimizer of the primitive variables form loss term \(\mathcal{L}_E\), that is
\begin{equation}
\theta^*=\mathop{\arg\min}_{\theta} r^T r
\end{equation}
The necessary condition for \(\theta^*\) to be a minimum point is \( \left( \nabla_{\theta^*} r \right)^T r =0 \). Considering that \(r\) is not necessarily 0, this implies  \(  \nabla_{\theta^*} r   =0\). Substituting this into the gradient term (\ref{grad}) of the variable-conservation form, we obtain
\begin{equation}
    \left(\nabla_{\theta^*} r \right)^T A^T A r=0
\end{equation}
However, the high-order small term \(r^T (\nabla_{\theta^*} A)^T Ar \) in the gradient of the variable-conservation form is generally non-zero (except in special cases). Therefore, \(\theta^*\) is not necessarily the minimizer of the variable-conservation form loss term \(\mathcal{L}_E\), i.e.,
\begin{equation}
    \theta^* \neq \mathop{\arg\min}_{\theta} r^T A^T Ar
\end{equation}
The gradient dynamics with respect to \(\theta\) for the two forms are given by
\begin{align}
    \frac{d \theta}{dt} &=-2\left(  \nabla_{\theta }r \right)^T r \\
    \frac{d \theta}{dt} &=-2\left(  Q^T \nabla_{\theta }r \right)^T \Lambda (Q^T r) - 2r^T (\nabla_{\theta } A )^T Ar
\end{align}
The primitive variables form simply superimposes each conservation law and performs gradient descent directly, whereas the variable-conservation form integrates the conservation laws and the physical properties at each point, performing gradient descent in the characteristic field. Notably, the local minima of the variable form are generally not those of the variable-conservation form, implying that the latter is less prone to getting trapped in local minima during optimization.

Taking the static water as an example, after the initial training period, the velocity components \(u\) and \(v\) are close to zero. The eigenvalues \(\lambda\) of matrix \(A\) are approximately \((1,h^2  ,h^2)\), and the eigenvector matrix Q is nearly the identity matrix. Therefore,
\begin{equation}
\left(  \nabla_{\theta }r \right)^T A^T A r=\left(\lambda \otimes \left(  \nabla_{\theta }r \right)^T \right)r
\end{equation}
It can be observed that at any given point, the greater the water depth \(h\), the larger the gradient weights for the momentum equations, and conversely, the smaller the water depth, the smaller the weights. Intuitively, a greater water depth results in a higher flow flux at the same velocity, leading to amplified errors in flow flux due to velocity inaccuracies. This disrupts both mass conservation and momentum conservation. Hence, it is necessary to increase the penalty on the momentum equation to mitigate velocity errors and ensure better conservation properties.

The imbalance among loss terms is one of the primary challenges in training PINNs. Matrix \(A\) enables pointwise automatic weighting of the loss term \(\mathcal{L}_E\) for the primitive variables form of SWE, and it is naturally derived from the conservation laws! This approach fundamentally differs from other techniques such as direct weighting of loss terms, gradient norm-based reweighting \citep{RN403}, and residual-based resampling \citep{RN404}. Based on this analysis, we conclude that the variable-conservation form of the shallow water equations is more suitable for embedding into the loss function of PINNs.

In addition, the above analysis provides valuable insights for constructing a self-adjusting weight matrix that is more suitable for specific simulation tasks based on matrix \(A\). For instance, to penalize high terrain change rates, we could construct a matrix \(A\) that incorporates terrain gradient information, such as
\begin{equation}
    A=\begin{bmatrix}
        1 &  z_y  & z_x\\
        u &  h    &  0 \\
        v &  0    &  h 
    \end{bmatrix}
\end{equation}
On the other hand, it is natural to question whether introducing additional conservation laws would yield better performance. For instance, if we incorporate the energy conservation equation, the matrix \(A\) can be rewritten as:
\begin{equation}
    A=\begin{bmatrix}
      1 & 0 & 0 &0\\
      u & h & 0 &0\\
      v & 0 & h &0\\
      (u^2+v^2)/2+g(h+z) & hu & hv &0
    \end{bmatrix}
\end{equation}
It can be observed that the energy conservation equation of SWE is not independent of the mass and momentum conservation equations. But its introduction will redistribute the weights.

\section{Settings for cases and models}
To verify the capability of PINNs in solving 2D SWE with rainfall and topography, we design a series of test cases. Specifically, we first set up three static water cases involving topography, followed by four static cases incorporating both rainfall and topography, and finally four dynamic cases. Detailed information for each case is presented in the respective subsections of this section. The computational results are discussed in Section \ref{sec4}.
\subsection{Static cases and well-balance} \label{sec3_1}
We first introduce three simple static water cases to evaluate the PINN's capability in handling terrain topologies.

The shallow water equations (\ref{swe}) without the rainfall source term and friction term, are also referred to as the balance laws because they require balancing water movement with terrain topology. In the context of numerical methods for balance laws, there exists a simple yet significant case: the steady state, particularly the static state where \(u=0\), \(v=0\) and \(h+z=const\). Although this problem is straightforward and allows for an analytical solution a priori, solving it numerically can be challenging due to the terrain topology. A numerical method is termed "well-balanced" if it can solve the steady-state of water. Developing "well-balanced" numerical methods remains a critical research issue \citep{RN381,RN384}.

When solving SWE using PINNs, the "well-balance" problem is also encountered. This problem requires that the PINN output remains time-invariant and that the predicted water depth aligns accurately with the terrain. The "well-balance" issue is a fundamental challenge in solving the shallow water equation with PINNs, as it determines whether PINNs can accurately represent terrain topology and subsequently simulate more complex dynamic scenarios. The simulation results presented in Section \ref{sec4_1} demonstrate that the PINN method can effectively address the "well-balance" challenge.

Considering the unevenness of real-world terrain, we set up three static water cases: Bump Topography, Depression Topography, and Tidal Topography, denoted as cases sB1, sD2, and sT3, respectively. The topological functions for these three cases are provided below. PINN requires the partial derivatives of each terrain topological function with respect to \(x\) and \(y\), which are not listed here for brevity.

The maximum height of the bump terrain case sB1 is 0.2 m, with a radius of 5 m. Its topological function is defined as
\begin{equation}
    z = 
\begin{cases} 
0.1 \left[ 1 + \cos\left(\frac{\pi}{25}(x^2 + y^2)\right) \right] & \text{if } x^2 + y^2 \leq 5^2  \\
0 & \text{elsewhere} 
\end{cases} 
,\quad \quad (x,y) \in [-10,10]^2
\end{equation}

The depression terrain case sD2 is the negative of the convex terrain case sB1, and its topological function is given by
\begin{equation}
    z = 
\begin{cases} 
-0.1 \left[ 1 + \cos\left(\frac{\pi}{25} (x^2 + y^2)\right) \right] & \text{if } x^2 + y^2 \leq 5^2  \\
0 & \text{elsewhere} 
\end{cases} 
,\quad \quad (x,y) \in [-10,10]^2
\end{equation}

The tidal terrain case sT3, derived from the "tide" problem used to validate the numerical method for 2D SWE \citep{RN406}, has the following topological function
\begin{equation}
z=0.2 \cos{\left(\frac{\pi}{10} x\right)}  \cos{\left(\frac{\pi}{10} y \right)}, \quad \quad (x,y)\in [-10,10]^2
\end{equation}

The water level for all three static water cases is uniformly set to 0.3 m, so the water depth is given by
\begin{equation}
    h=0.3-z
\end{equation}

Let \(B\) denote the four boundaries of the domain \([-10,10]^2\). For all three static water cases, Dirichlet boundary conditions are applied, such that
\begin{equation}
    u(x,y,t)=v(x,y,t)=0, \quad \quad     (x,y)\in B
\end{equation}

The computational time interval for these three cases is from 0 to 5 seconds. Note that no boundary conditions are specified for the water depth \(h\). It is expected that PINNs will automatically solve for the water depth based on the velocity boundary conditions and initial conditions.

\subsection{Static cases with rain} \label{sec3_2}
Rainfall, as a critical factor in flash floods and urban flooding, has not been considered in current research on solving SWE using PINNs. Therefore, this study investigates the ability of PINNs to handle SWE with rainfall source terms, aiming to facilitate the practical engineering application of PINNs. In this work, the rainfall scenario is based on the 100-year design rainfall intensity for a city in China and is transformed into the Chicago rainfall pattern. The design rainfall duration is 5 minutes, with a cumulative rainfall of approximately 24.02 mm. The peak coefficient of the Chicago rainfall pattern \citep{RN368} is 0.5, and the formula for rainfall intensity (unit: mm/min) is given by
\begin{equation}
    rain=73.78 \frac{(t-2.5)/0.5 \times 0.13+18.1}{(t+18.1)^{1.87} } 
\end{equation}

We first eliminate the terrain factor by setting up a flat terrain case sFR4 to independently examine the ability of PINNs to solve SWE with only the rainfall source term. Subsequently, we integrates the rainfall source term and terrain to investigate the effectiveness of PINNs in addressing the "well-balance" problem of static water under rainfall conditions. Specifically, the three terrain topologies from Section \ref{sec3_1} are combined with the rainfall scenario designed in this section to form cases sBR5, sDR6, and sTR7. For these four cases, no boundary conditions are imposed on water depth, while Dirichlet boundary conditions are applied to velocity, i.e.,\( u(x,y,t)=v(x,y,t)=0\) for \((x,y)\in B\). The computational time interval is from 0 to 5 minutes. Furthermore, these four cases utilize the dimensional transformation technique and the entropy stability condition.

\subsection{Dynamic cases}
In dynamic problems, the Riemann problem is a fundamental issue for hyperbolic conservation laws and serves as the cornerstone of modern computational fluid dynamics. A typical example of the Riemann problem in the context of SWE is the dam break problem. We consider the Riemann problem as a foundational issue for evaluating the capabilities of PINNs, and thus begins by solving the dam break problem in dynamic scenarios. The pseudo-2D dam break case dDB8 extends the one-dimensional dam break experiment to two dimensions. Dazzi et al. \citep{RN378} demonstrated that PINNs can effectively solve 1D dam break problem. While many 1D numerical methods can be extended to two-dimensional cases, for PINNs, the transition from one to two dimensions presents a nearly new challenge due to the increased number of variables, more complex nonlinearities, and additional imbalance loss terms. The pseudo-2D dam break case dDB8 has a flat terrain, and the initial conditions are as follows
\begin{equation}
    h(x,y,0) = 
\begin{cases} 
2, & \text{if } x \leq 0 \\
1, & \text{if } x > 0 
\end{cases}, \quad 
u(x,y,0) = v(x,y,0) = 0, \quad (x,y) \in [-10,10]^2.
\end{equation}

Regarding the boundary conditions, the water depth in the \(x\)-direction is subject to Dirichlet boundary conditions, while the remaining boundaries have Neumann conditions with zero gradients. The case dDB8 is a Riemann problem with an analytical solution. We employ the open-source algorithm code for the Riemann problem provided by Dazzi et al. (Dazzi 2024) to obtain the accurate solution of this problem, which serves as a benchmark for comparing with PINN results.

The circular dam break case dCDB9 simulates the flow phenomenon of a cylindrical water body breaking and spreading radially outward, primarily characterized by an internal rarefaction wave and an external shock wave. The experiment assumes no topographic changes. The initial conditions are as follows
\begin{equation}
    h(x,y,0) = 
\begin{cases} 
2, & \text{if } x^2 + y^2 \leq 3^2 \\
1, & \text{elsewhere}
\end{cases}, \quad
u(x,y,0) = v(x,y,0) = 0, \quad (x,y) \in [-10,10]^2
\end{equation}

The tidal problem dT10 simulates water flow over uneven terrain to evaluate PINN's capability in handling flow phenomena similar to actual floods. The tidal problem with rainfall case dTR11 models tidal flows under rainfall conditions, representing a scaled-down version of real-world rainfall and flooding. The rainfall data are taken from the designed rainfall in Section \ref{sec3_2}. Based on the definition of the tidal problem by Supei et al. \citep{RN406}, the topographic function and initial conditions for this study's tidal problem are set as follows
\begin{align}
    &z=1+0.01 \cos{\frac{\pi}{2} x} \cos{\frac{\pi}{2} y}\\
    &u(x,y,0)=v(x,y,0)=0 \\
    &h(x,y,0)=z 
\end{align}
where \((x,y)\in [-2,2]^2\).

Cases dCDB9, dT10, and dTR11 currently lack analytical solutions. In this study, the HLL scheme is employed to numerically solve dCDB9, while the entropy stable scheme ES1 is used to solve dT10 and dTR11, serving as benchmarks for comparison with PINNs. The HLL scheme simplifies the solution of the Riemann problem by neglecting the contact wave, offering advantages in computational complexity \citep{RN385}. The ES1 scheme satisfies the discrete entropy inequality, ensuring energy stability and being "well-balanced". For more detailed information, please refer to this literature \citep{RN384}. Both algorithms are implemented using GPU parallel computing; further details can be found in the open-source code provided in this study.

\subsection{Models setting}
We construct the PINN using a fully connected neural network with inputs \(x,y,t\) and outputs \(h,u,v\). The general settings are as follows. The network consists of five hidden layers, each containing 300 neurons, and the activation function is the hyperbolic tangent (tanh). To enhance the network's adaptability and training efficiency, we employ the Neuron-wise Local Adaptive Activation Function (N-LAAF) technique, which introduces a scalable parameter for each neuron to achieve local adaptation of the activation function. Additionally, N-LAAF incorporates a gradient recovery term in the loss function to accelerate convergence \citep{RN377}. The optimizer is Adam, with an initial learning rate of 1e-3 and a decay factor of 1e-2. Parameters are initialized using Glorot normal initialization. Training points are randomly sampled from the spatiotemporal domain, including 8192*4 internal points, 2048*4 boundary points, and 2048*4 initial points. For the loss function, weights are assigned as follows: all boundary conditions, regardless of type, have a weight of 1; all initial conditions have a weight of 100; non-negative water depth and dry-wet conditions have a weight of 10. The weights for differential equations vary depending on the specific case. Where applicable, deviations from these general settings will be detailed separately.

For the static water cases sB1, sD2, and sT3, the total number of training steps is 50,000, with the learning rate decaying at step 30,000. The weights for mass conservation, x-momentum conservation, y-momentum conservation in the differential equations, and entropy conservation condition are set to 1, 10, 10, and 1, respectively.

In the dynamic pseudo-2D dam break case dDB8, to facilitate the comparison of various forms of shallow water equations, the weights for the differential equation terms and entropy stability conditions are both set to 1. The total number of training steps is 30,000, with the learning rate decaying at step 20,000. The hidden layer configuration is 4*300 neurons, which provides slightly better performance compared to the 5*300 configuration.

In the dynamic circular dam break case dCDB9, the total number of training steps is 5,000 with no learning rate decay. For the tidal problem case dT10, the total number of training steps is 100,000, and the learning rate is reduced every 20,000 steps. In the tidal problem with rainfall case dTR11, the total number of training steps is 120,000, and the learning rate is reduced every 50,000 steps. All three cases employ the variable-conservation form of SWE, with loss function weights set to 10, 1, and 1, respectively.

Why are such large weights assigned to the initial conditions? Firstly, the solution space of partial differential equations (PDEs) can contain an infinite number of solutions. Therefore, initial and boundary conditions play a critical role in determining the unique solution. Specifically, for hyperbolic PDEs like the shallow water equations, the initial conditions propagate within the characteristic domain. Any deviation in the initial conditions may lead to significant distortions as they propagate, resulting in solutions that deviate from the true solution. Secondly, according to Wang et al. \citep{RN375}, the gradient of the boundary condition residual term tends to be more concentrated around zero compared to the equation residual term, indicating that the magnitude of the gradient for the boundary condition residual is smaller. This gradient imbalance necessitates the introduction of additional weight parameters to ensure balanced optimization. Given the use of a square loss function, as the residuals of the boundary and initial conditions decrease, their gradients diminish, potentially leading them to exit the primary optimization focus. In this study, the boundary conditions are relatively simple, hence the assigned weights are smaller.

In this study, the Mean Absolute Error (MAE) and Root Mean Square Error (RMSE) are employed to quantify the discrepancy between the PINN solution \(U_{\theta}\) and the true or numerical solution. The definitions of MAE and RMSE are as follows, where \(N\) represents the number of comparison points
\begin{align}
    \text{MAE}  &= \frac{1}{N} \sum_{i=1}^{N} \left| U_{\theta,i} - U_i \right| \\
\text{RMSE} &= \left( \frac{1}{N} \sum_{i=1}^{N} \left| U_{\theta,i} - U_i \right|^2 \right)^{1/2}
\end{align}

\section{Results and Discussion} \label{sec4}

\subsection{Results of Static Water Cases} \label{sec4_1}

\begin{figure}
    \centering
    \includegraphics[width=1\linewidth]{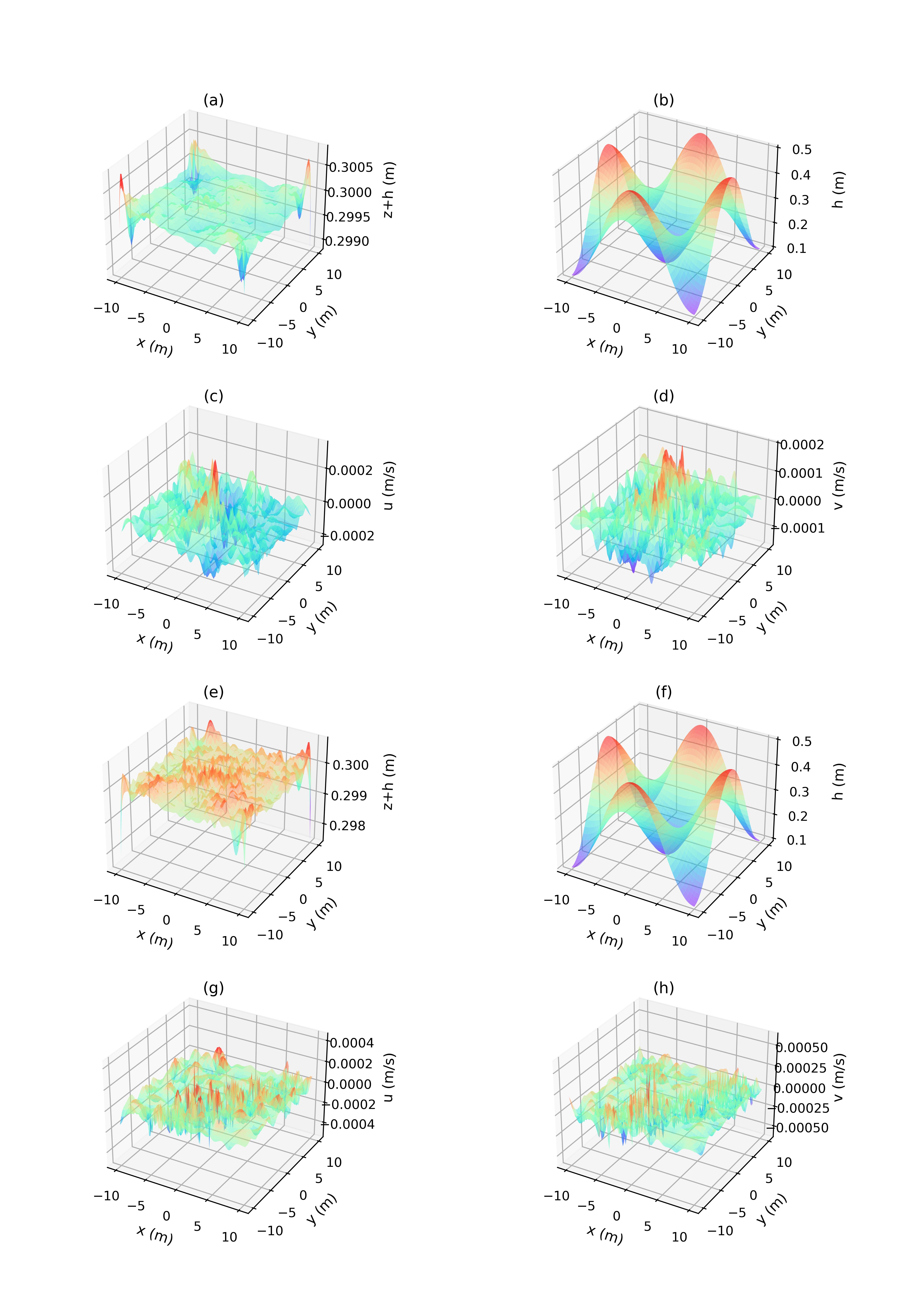}
    \caption{The static tidal case sT3. (a) and (e) water level, (b) and (f) water depth, (c) and (g) flow velocity \(u\), and (d) and (h) flow velocity \(v\), at the 5th second. (a)-(d) is with entropy conservation condition, and (e)-(f) is without entropy conservation condition.}
    \label{fig1}
\end{figure}

First, we analyse the static water cases in Section \ref{sec3_1} to evaluate the ability of the PINN to handle "well-balance" problems and compare the effects with and without the entropy conservation condition. All three static water cases employ the variable-conservation form of SWE. The MAE and RMSE between the approximate solutions \(U_\theta\) and the exact solutions at the final time for cases sB1, sD2, and sT3 are presented in Table \ref{tab1}. Regardless of whether the entropy conservation condition is applied, the errors remain within the range of 1E-4m to 1E-5m, indicating a high level of approximation accuracy. It is observed that the approximation accuracy is comparable in both scenarios, with neither one demonstrating a significant advantage. For brevity, we focus on the most complex terrain case, sT3. The approximate solutions for water level, water depth, and flow velocity at the 5th second for case sT3 are illustrated in Figure \ref{fig1}, where (a-b-c-d) includes the entropy conservation condition and (e-f-g-h) does not. All figures are generated from the trained model's approximate solutions at uniform grid points, which serve as test points with a grid spacing of 0.2m. The results show that the water level \(h+z\) remains stable around 0.3m, and the flow velocities \(u\) and \(v\) are close to 0m/s, aligning well with the true values. The water depth also closely matches the topography. Comparing the two approaches, it is evident that the entropy conservation condition helps mitigate micro-local oscillations, but this benefit is minimal and does not justify the additional computational cost.

In summary, the PINN used in this study effectively handles "well-balance" problems and demonstrates the capability to manage 2D terrain topologies. In the static water cases, the approximation accuracy is consistent regardless of whether the entropy conservation condition is applied. Notably, both with and without the entropy conservation condition, the approximate solutions \(U_\theta\) exhibit significant micro-local oscillations, particularly in regions where the terrain topology changes.

\begin{table}
\caption{Static water cases sB1, sD2, and sT3. Errors between the approximate solutions \(U_\theta\) and the exact solutions at the 5th second. The units of \(h\), \(u\) and \(v\) are m, m/s and m/s respectively.}
\label{tab1}
\centering
\begin{tabular}{l c c c c c c}
\toprule
& \multicolumn{6}{c}{with entropy conservation condition}\\ 
\cline{2-7}
{Cases} &\multicolumn{3}{c}{MAE}  & \multicolumn{3}{c}{RMSE} \\
\cline{2-7}
&\(h\) &\(u\)  &\(v\)  &\(h\) &\(u\)  &\(v\)\\
\midrule
sB1 &8.4E-05 &1.1E-04 &1.1E-04  &1.5E-04  &3.1E-04 &4.0E-04 
\\
\hline
sD2	& 9.6E-05 &2.3E-05	&2.5E-05 &1.2E-04 &3.5E-05 &3.9E-05	
    \\
\hline
sT3	& 7.5E-05 &4.7E-05	&3.0E-05 &1.0E-04 &6.3E-05 &3.9E-05	
    \\
\hline
\hline
& \multicolumn{6}{c}{without entropy conservation condition} \\
\hline
sB1 &9.2E-05 &5.1E-05 &6.8E-05  &1.1E-04  &7.0E-05 &8.6E-05\\
\hline
sD2 &4.5E-05  &2.9E-05	&3.1E-05 &9.3E-05 &9.1E-05 &1.4E-04 \\
\hline
sT3	&1.2E-04  &4.1E-05	&4.5E-05 &1.7E-04 &5.7E-05 &6.3E-05\\
\bottomrule
\end{tabular}
\end{table}

\begin{figure}
    \centering
    \includegraphics[width=1\linewidth]{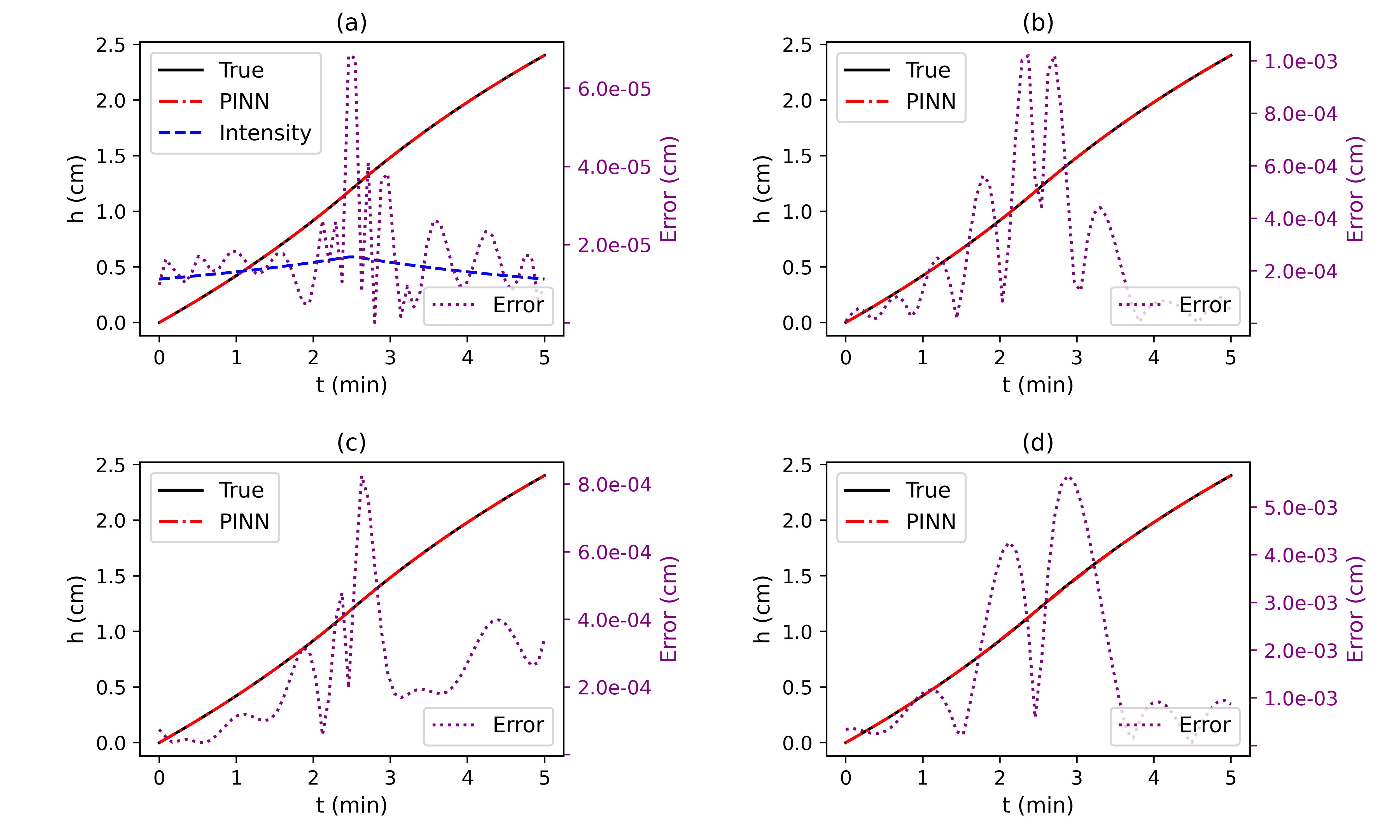}
    \caption{Comparison of cumulative water depth growth and cumulative rainfall for (a) sFR4, (b) sBR5, (c) sDR6, and (d) sTR7. The black solid line represents the cumulative rainfall depth from the designed rainfall, the red dashed line denotes the approximate solution \(U_\theta\), the purple dotted line indicates the point-wise absolute error between them, and in panel (a), the blue dashed line represents the rainfall intensity.}
    \label{fig2}
\end{figure}

\subsection{ Results of static water cases with rain}
This section investigates the capability of PINN to solve SWE with rainfall source term.

All four cases utilize the variable-conservation form of SWE with entropy stable condition and undergo the dimensional transformation. After model training, uniform grid points are selected as test points with a spatial resolution of 0.2m and a temporal interval of 1 minute. For the four cases, the comprehensive error analysis is summarized in Table \ref{tab2}, which presents the MAE and RMSE between the approximate solution \(U_\theta\) and the true solution at both the initial and final moments. The water depth errors range from 1E-3 cm to 1E-5 cm, while the velocity errors range from 1E-4 cm/min to 1E-5 cm/min, indicating high approximation accuracy.

The cumulative growth process of water depth and its error for the four static water cases with rainfall are illustrated in Figure 2. Notably, the error axis on the right side of each figure spans from 1E-3 cm to 1E-5 cm, indicating that the point-wise absolute error between the approximate solution \(U_\theta\) and the true solution is relatively small. Observing the error plots, a significant feature emerges: the error does not accumulate over time but instead exhibits a distribution pattern that is higher in the middle and lower at the edges. This characteristic can be attributed to the Chicago rain pattern, where rainfall intensity initially increases and then decreases, resulting in a high-in-the-middle, low-on-the-sides distribution. This behaviour contrasts with traditional numerical methods, where errors typically accumulate over time steps.

In conclusion, the PINN employed in this study demonstrates the capability to solve the SWE with rainfall source term, achieving high accuracy that meets the practical requirements for rainfall flood simulation. It is important to note that PINN exhibits sensitivity to rainfall intensity; specifically, greater intensities result in larger local errors.

\begin{table}
\caption{The four static water cases with rain (sFR5, sBR5, sDR6, and sTR7) and the three dynamic cases (dCDB9, dT10 and dTR11). Errors between \(U_\theta\) and the true solution at the initial and final moments of the three dynamic cases. (The units of \(h\), \(u\) and \(v\) are cm, cm/min and cm/min respectively for static cases. The units of \(h\), \(u\) and \(v\) are m, m/s and m/s respectively for dynamic case.)}
\label{tab2}
    \centering
    \begin{tabular}{l c c c c c c}
    \toprule
    & \multicolumn{6}{c}{ the initial moment}\\ 
    \cline{2-7}
    {Cases} &\multicolumn{3}{c}{MAE}  & \multicolumn{3}{c}{RMSE} \\
    \cline{2-7}
    &\(h\) &\(u\)  &\(v\)  &\(h\) &\(u\)  &\(v\)\\
    \midrule
    sFR4&1.7E-05	&1.1E-05	&1.3E-05	&6.9E-05	&2.7E-05	&2.1E-05\\
    \hline
    sBR5&3.1E-04	&9.1E-05	&2.6E-04	&4.9E-04	&2.3E-04	&3.0E-04	\\
    \hline
    sDR6&3.4E-04	&8.0E-05	&1.4E-04	&6.3E-04	&1.7E-04	&2.2E-04	\\
    \hline
    sTR7&1.8E-03	&1.2E-04	&1.3E-04	&2.3E-03	&2.5E-04	&3.2E-04	\\
    \hline
    dCDB9 &5.9E-03 &8.3E-03	&6.4E-03	&3.2E-02	&1.0E-02	&7.7E-03	\\
    \hline
    dT10 &7.0E-05	&4.6E-05	&5.9E-05	&8.7E-05	&6.4E-05	&8.0E-05	\\
    \hline
    dTR11   &1.3E-05	&1.3E-05	&1.2E-05	&1.7E-05	&1.8E-05	&1.7E-05	\\
    \hline
    \hline
    & \multicolumn{6}{c}{the final moment} \\
    \midrule
    sFR4&1.7E-04	&9.8E-05	&1.1E-04	&2.4E-04	&1.4E-04	&1.5E-04 \\
    \hline
    sBR5&1.4E-03	&3.2E-04	&3.6E-04	&1.9E-03	&6.4E-04	&5.8E-04\\
    \hline
    sDR6&1.1E-03	&8.8E-05	&1.4E-04	&2.0E-03	&2.4E-04	&2.8E-04\\
    \hline
    sTR7&4.3E-03	&2.0E-04	&3.4E-04	&5.6E-03	&5.1E-04	&9.5E-04\\
    \hline
    dCDB9&5.6E-03	&1.7E-02	&1.3E-02	&1.0E-02	&2.8E-02	&2.7E-02 \\
    \hline
    dT10&2.6E-04	&4.9E-04	&4.8E-04	&3.9E-04	&6.6E-04	&6.2E-04\\
    \hline
    dTR11&2.5E-04	&1.5E-04	&1.5E-04	&3.0E-04	&1.9E-04	&1.9E-04\\
    \bottomrule
    \end{tabular}
\end{table}

\subsection{Dynamic cases}
This section aims to evaluate the capability of the PINN used in this study to handle dynamic problems. The Riemann problem serves as a fundamental benchmark for assessing PINN's performance, but its two-dimensional case has not yet been verified. For PINN, one-dimensional experience cannot be directly extrapolated to two dimensions, as the latter presents a more complex optimization challenge. Therefore, this section first investigates two two-dimensional dam break problems, dDB8 and dCDB9, followed by an exploration of the tidal problem dT10 and the tidal problem with rainfall dTR11.

\begin{figure}
    \centering
    \includegraphics[width=1\linewidth]{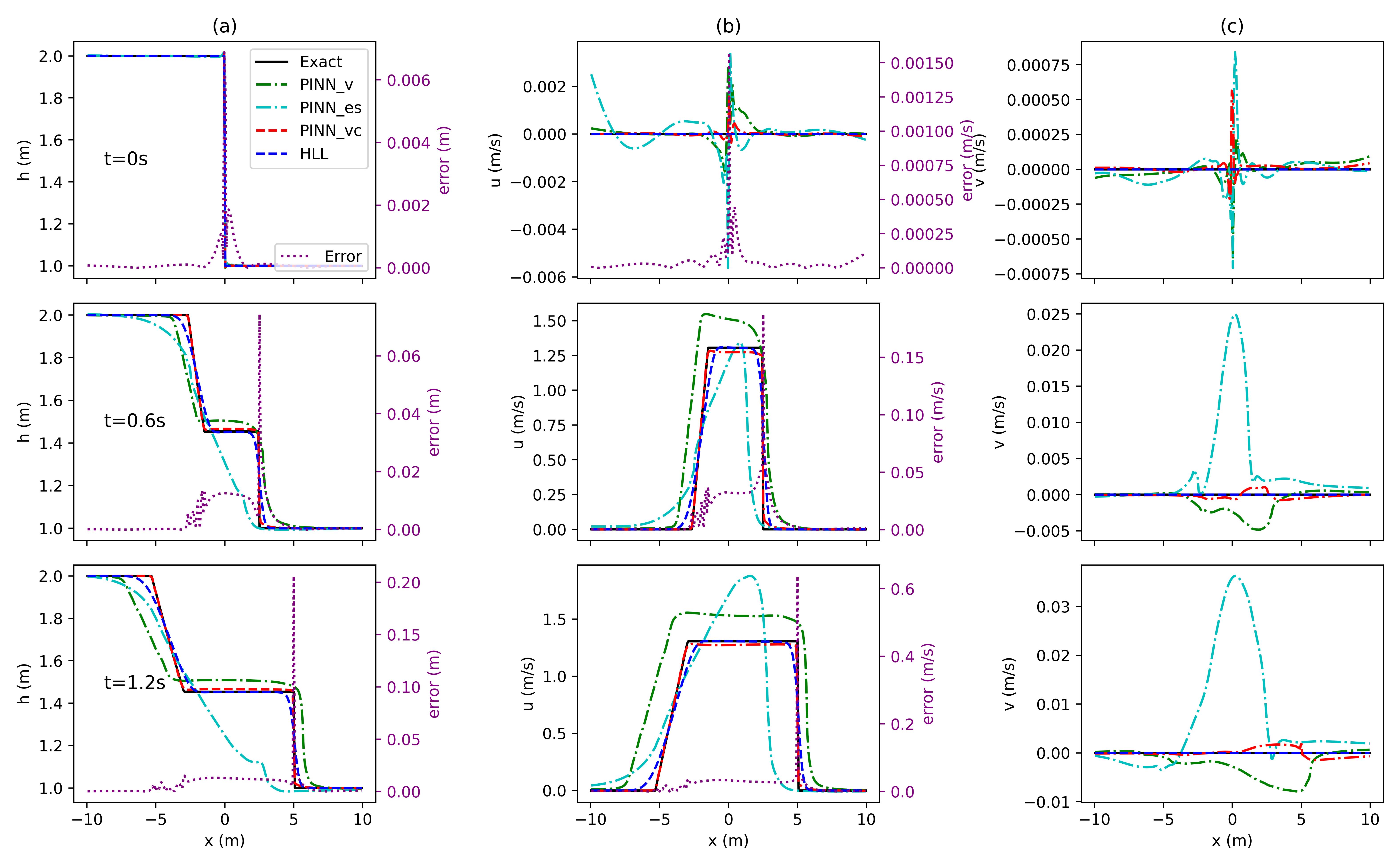}
    \caption{The pseudo 2D dam break case dDB8. This figure compares the approximate solutions of PINN with different equation forms, the HLL numerical solution, and the exact solution. Specifically, PINN\_v represents the primitive variables form PINN, PINN\_es denotes the variable-conservation form PINN with an entropy stability condition, and PINN\_vc is the variable-conservation form PINN. Columns (a), (b), and (c) correspond to water depth, velocity \(u\), and velocity \(v\), respectively. The secondary y-axis on the right side of columns (a) and (b) displays the pointwise absolute error between PINN\_vc and the exact solution.}
    \label{fig3}
\end{figure}

First is the pseudo 2D dam break case dDB8. As shown by the green dotted line in Figure \ref{fig3}, the primitive variables form PINN captures the movement trends of the shock wave and rarefaction wave but exhibits significant errors in velocity values, leading to substantial discrepancies in water depth. The variable-conservation form PINN with the entropy stability condition fails to solve this dam break case effectively, as indicated by the light green dotted line in Figure \ref{fig3}. Specifically, the rarefaction wave over-propagates, failing to maintain the correct water level between the rarefaction and shock waves, and only a minor shock wave is present.

Observing the error curve of the variable-conservation form PINN, it is evident that it closely approximates the exact solution almost everywhere, demonstrating very high accuracy. The HLL numerical solution serves as a reference standard, indicating that the approximate solution of the variable-conservation form PINN generally outperforms the first-order accuracy solution of the HLL scheme with a grid size of 0.08m and a CFL number of 0.25. From this, we can conclude that PINN has the capability to obtain high-precision solutions for the 2D SWE. However, it is important to note that we cannot claim PINN is superior to traditional numerical schemes.

Comparing the approximate solutions of the primitive variables form PINN and the variable-conservation form PINN, both trained with identical parameters, reveals that the variable-conservation form PINN achieves significantly higher approximation accuracy. Combining this observation with the theoretical analysis of equation forms presented in Section \ref{sec2_3}, it can be concluded that embedding the variable-conservation form of SWE into the PINN loss term may be a more optimal choice.

Surprisingly, the variable-conservation form PINN with entropy condition performed poorly in the dam break problem. The addition of entropy condition not only failed to provide significant benefits but also over-smoothed the solution. While further extensive verification is necessary, the current results suggest that the variable-conservation form PINN without entropy condition is a more reliable option.

\begin{figure}
    \centering
    \includegraphics[width=1\linewidth]{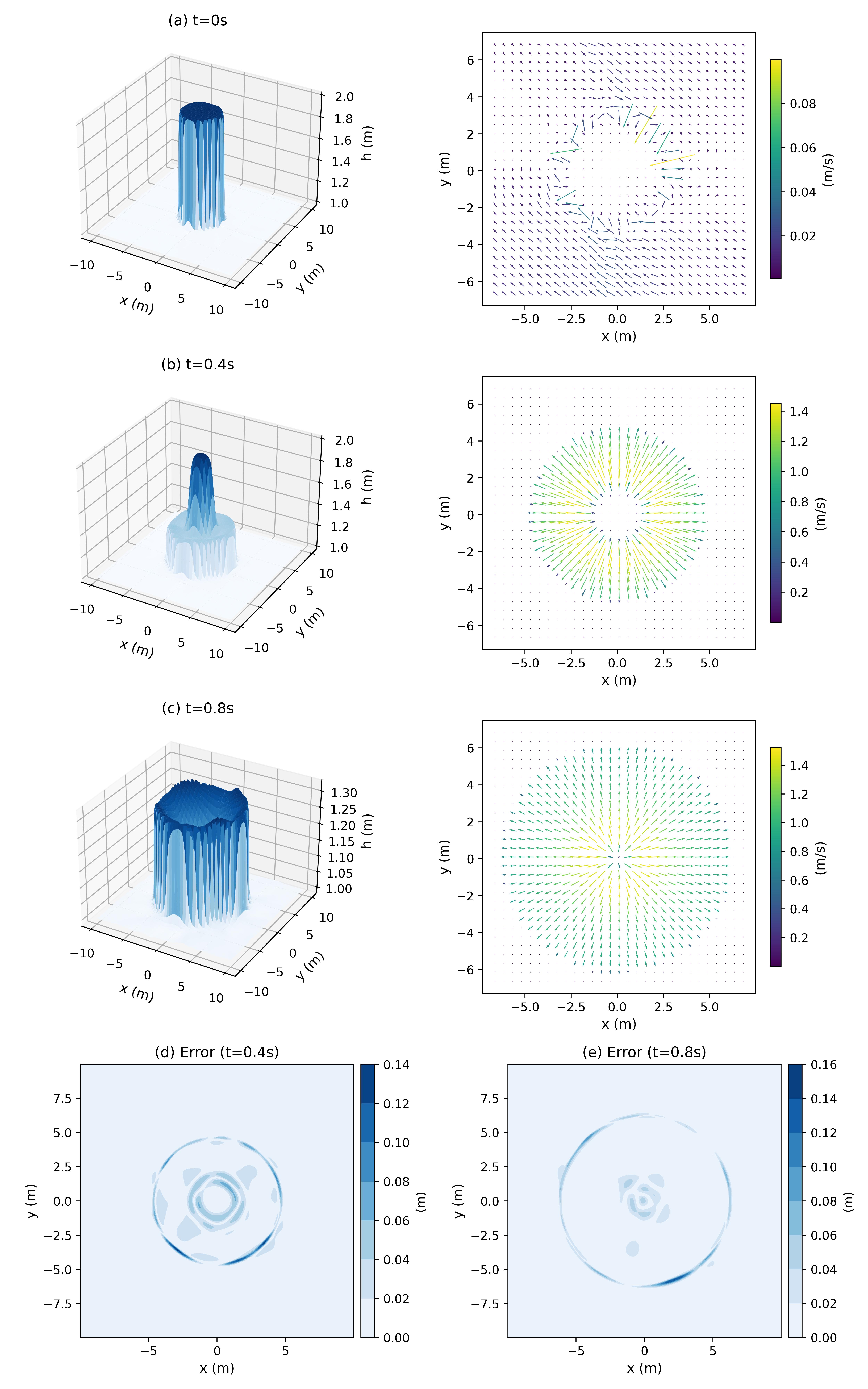}
    \caption{2D circular dam-break case dCDB9. Water depth and velocity at 0.0s, 0.4s, and 0.8s. And the error between the PINN approximate solutions and the HLL numerical solutions of water depth at 0.4s and 0.8s.}
    \label{fig4}
\end{figure}

The next challenging case is the 2D circular dam break case dCDB9. This problem generates more intricate flow patterns. During the initial stages of the dam break, circular shock waves propagate outward while circular rarefaction waves move inward. Upon reaching the center of the domain, the rarefaction wave causes a collapse in water depth towards the center, leading to significant water level reductions near the central region. The second row of Figure \ref{fig4} illustrates the flow field snapshot of the PINN approximate solution at 0.4 seconds, effectively capturing the structures of both the circular shock and rarefaction waves. The third row of Figure \ref{fig4} shows the flow field at 0.8 seconds, highlighting the steep shock wave structure and the central collapse. However, it is observed that the water depth at the shock front exhibits uneven distribution along the circumference. This phenomenon is attributed to the initial condition approximation errors, as depicted in the first row of Figure \ref{fig4}. Given the continuous yet limited approximation capabilities of the PINN function, deviations from strong discontinuous initial conditions are inevitable. These deviations propagate within the characteristic field, progressively increasing the discrepancy between the approximate and exact solutions.

To quantify the error of the PINN approximate solution, this study employs a first-order accurate HLL scheme for numerical simulation. The computational grid size is set to 0.04m*0.04m with a CFL number of 0.25, ensuring that the numerical solution closely approximates the exact solution. Errors are predominantly concentrated at the shock and rarefaction wave fronts, with high-error regions primarily located at the shock wave, as depicted in the fourth row of Figure \ref{fig4}. Additionally, the errors exhibit uneven distribution along the circumference, with high-error positions at both time instances largely corresponding, indicating that the initial condition deviations have disrupted symmetry and continue to propagate. The MAE and RMSE between the PINN approximate solutions \(U_\theta\) and the exact solutions at the initial and final times are summarized in Table \ref{tab2}. It is observed that the error magnitudes at the final time are consistently greater than or equal to those at the initial time.

In conclusion, PINN demonstrates effective performance in solving the 2D circular dam break problem. This study highlights that the accuracy of PINN's initial condition approximation fundamentally influences its overall accuracy in solving SWE. These findings underscore the critical importance of assigning appropriate weight to the initial conditions.
\begin{figure}
    \centering
    \includegraphics[width=1\linewidth]{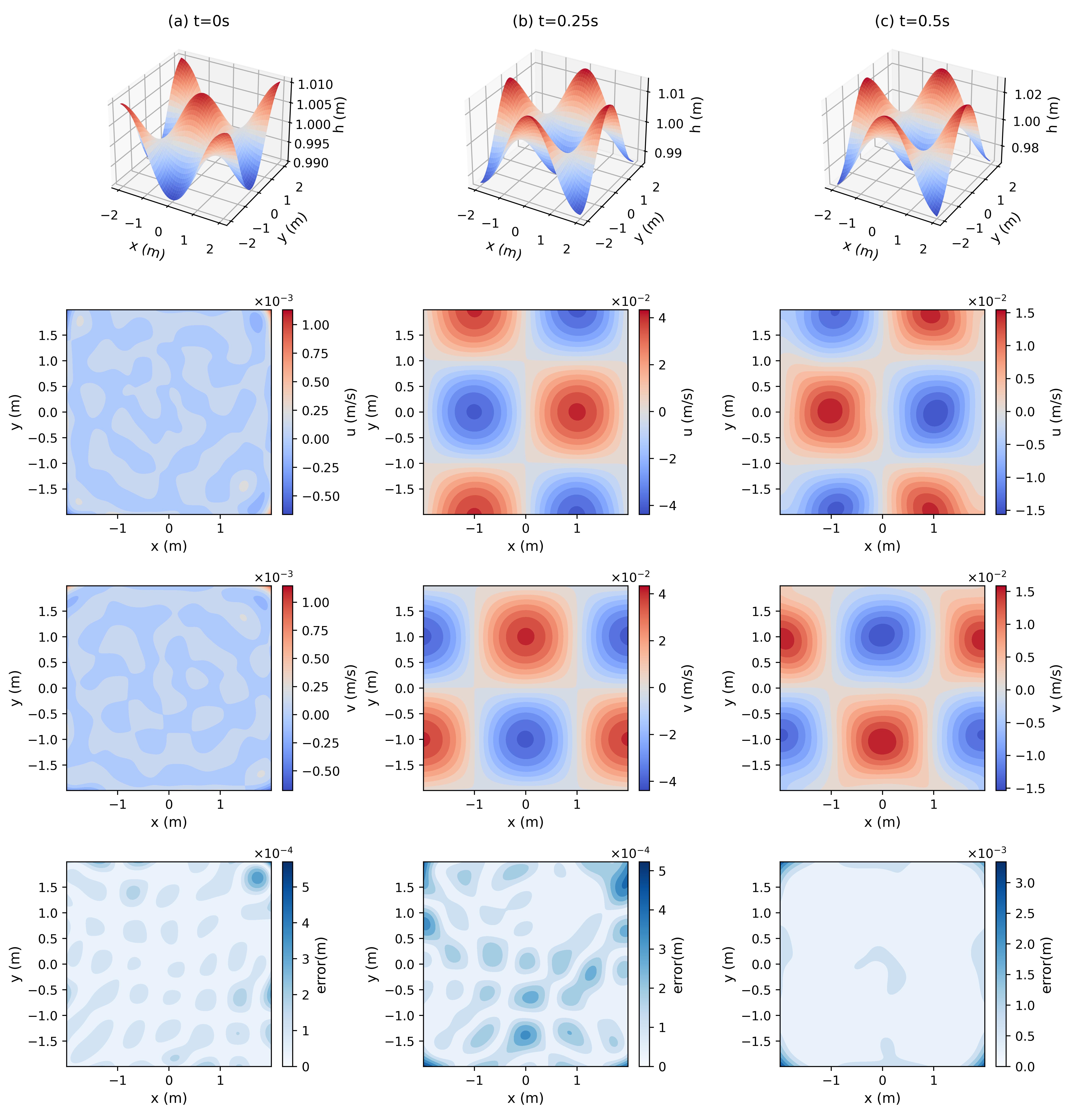}
    \caption{Tidal problem case dT10. Water depth, velocity \(u\) , velocity \(v\) and error at 0s, 0.25s, and 0.5s.}
    \label{fig5}
\end{figure}

\begin{figure}
    \centering
    \includegraphics[width=1\linewidth]{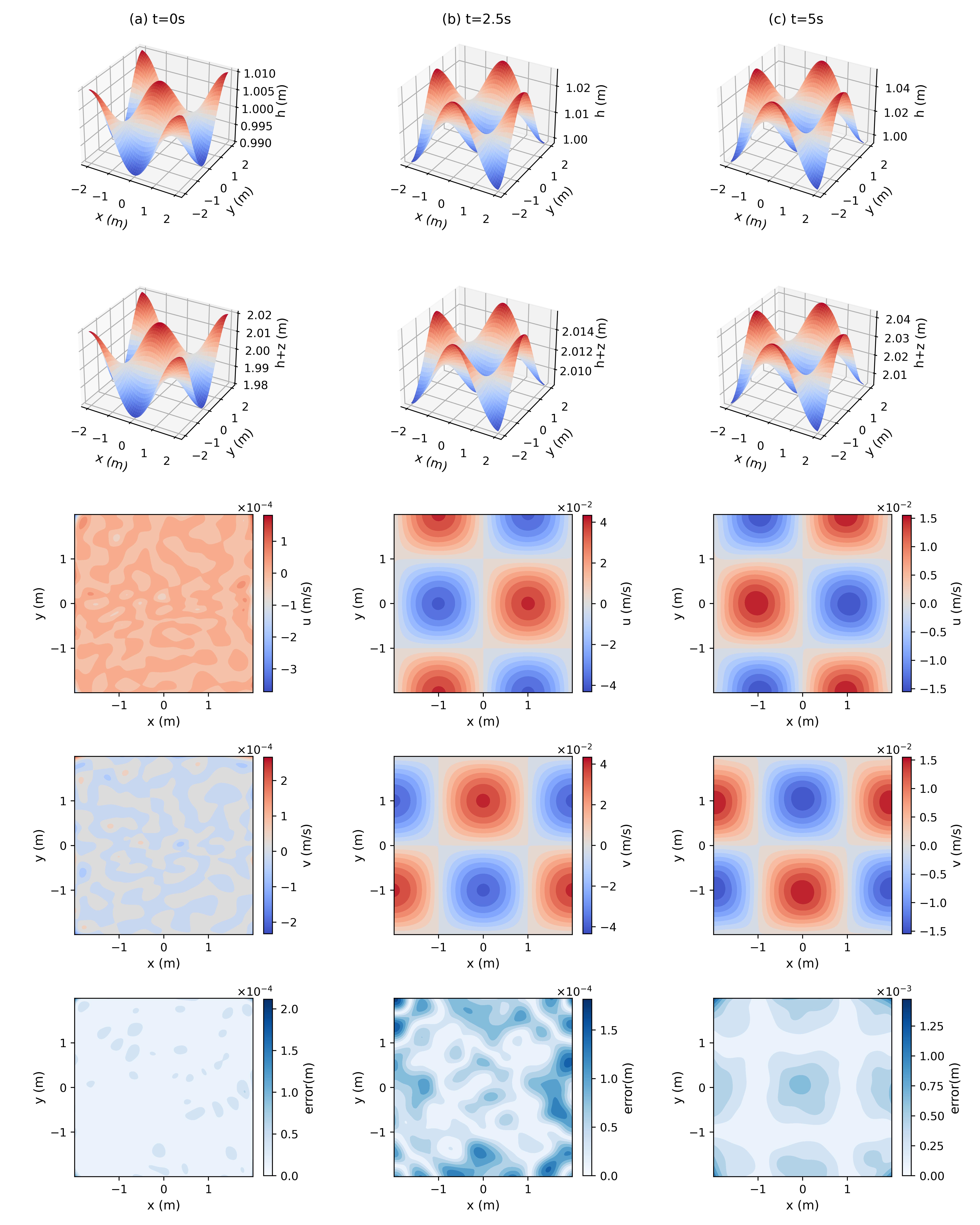}
    \caption{Tidal problem case dTR11. Water depth, water level, velocity \(u\), velocity \(v\) and error at 0s, 0.25s, and 0.5s.}
    \label{fig6}
\end{figure}
The following is a more challenging example of the variable topology tidal problem case dT10. The initial water depth for this problem is illustrated in Figure \ref{fig5} at 0s, and its topography function shares the same shape as the initial water depth. The initial water surface exhibits a pattern of higher levels in the centre and corners, with lower levels along the four sides. Consequently, under the influence of gravity, water flows toward the lower regions on the sides. The second column of Figure \ref{fig5} captures the instantaneous flow field at 0.25s, showing water converging towards the low areas on the four sides. At this moment, the water depth at the sides has gradually increased, while the depths in the centre and corners continue to decrease, as evidenced by the velocity components \(u\) and \(v\), which indicate ongoing flow towards the lower regions. Periodic boundary conditions and the geometric symmetry of the terrain result in the formation of a Riemann problem at the boundaries. Specifically, this means that the water depth remains constant while the velocities are equal in magnitude but opposite in direction, leading to the generation of waves that propagate in opposite directions. As shown in the rightmost figure of Figure \ref{fig5} at 0.5s, the flow direction has reversed compared to 0.25s, although the flow velocity magnitude is only about half of its previous value.

To calculate the error in the PINN approximate solution, we employ the first-order accurate entropy stable scheme ES1 proposed by Fjordholm et al. \citep{RN384} to numerically solve this variable topology tidal problem. This study employs a high-resolution grid with a spatial resolution of 0.01m*0.01m and a relatively low CFL number of 0.25, ensuring that the numerical solution is sufficiently accurate to approximate the exact solution.

The pointwise absolute errors between the PINN approximate solution and the ES1 numerical solution are illustrated in the fourth row of Figure \ref{fig5}. At 0.25s, the errors are relatively small but unevenly distributed. By 0.5s, the errors become slightly larger and are primarily concentrated at the four corners. The MAE and RMSE values for the initial and final moments are summarized in Table \ref{tab2}. All errors remain below the 1E-04m level, indicating a generally high approximation accuracy.

The water depth distribution and flow velocity contour maps obtained by PINN for solving the variable topology tidal problem exhibit excellent symmetry and clear structure. The results are in close agreement with the high-precision numerical solution. In summary, PINN demonstrates the capability to effectively solve the dynamic flow problems of 2D SWE with variable topologies, achieving satisfactory solution accuracy.

The final case is the tidal problem with rainfall dTR11. To facilitate comparison with the results of the tidal problem, this study compresses the duration of rainfall from 5 minutes to 0.5 seconds while maintaining the total rainfall volume constant, thereby increasing the rainfall intensity source term in Equation (\ref{swe}). The snapshots of water depth, water level, and velocity for the case at 0s, 0.25s, and 0.5s are presented in Figure \ref{fig6}. Over time, the water level consistently rises. Compared with Figure \ref{fig5}, the fundamental flow trend remains unchanged; water still converges towards lower areas.

To compute the errors, this case employs the ES1 numerical scheme with the same grid resolution and CFL number as used in case dT10. The pointwise absolute errors between the PINN approximate solution and the ES1 numerical solution at 0s, 0.25s and 0.5s are shown in Figure \ref{fig6}, while the MAE and RMSE values at the initial and final moments are provided in Table \ref{tab2}. Similar to the error distribution observed in case dT10, the errors at 0.5s are relatively larger, with high-error regions primarily concentrated at the center and the four corners and edges. The pointwise approximation errors remain below 1E-3m, and the overall average approximation error is below 1E-4m, indicating satisfactory approximation accuracy.

Initially, the total water volume is 16\(m^3\). At 0.25s and 0.5s, the total water volume in the PINN approximate solution is approximately 16.192m³ and 16.384\(m^3\), respectively. Dividing by the area of 16\(m^2\), the water level increases by 12mm and 24mm, respectively, which aligns with the cumulative rainfall. Therefore, the PINN approximate solution maintains water volume conservation in this dynamic case.

In conclusion, PINNs demonstrate the capability to effectively solve two-dimensional shallow water equations with topography and rainfall, with satisfactory solution accuracy.

\subsection{Discussion}
Based on the comprehensive analysis of all cases, it can be concluded that PINNs are capable of effectively solving two-dimensional shallow water equations with topography. For numerical methods addressing topographic problems, a carefully designed "well-balanced" numerical scheme is essential. In contrast, PINNs can achieve satisfactory results with a simple fully connected neural network structure. Observations indicate that the error in the PINN approximation solution is positively correlated with the terrain slope; regions with steeper slopes exhibit larger approximation errors. Moreover, steeper slopes also increase the training difficulty. It is important to note that this challenge is not unique to PINNs; high-slope or highly variable terrains pose similar difficulties for traditional numerical methods. To mitigate this issue in PINNs, techniques such as slope-based resampling, slope-based reweighting, or embedding slope information into the equation, as proposed in Section \ref{sec2_3}, could be employed. These approaches are conceptually analogous to mesh refinement techniques used in numerical methods.

By integrating both static and dynamic rainfall cases, it can be concluded that PINNs are capable of effectively solving the two-dimensional shallow water equations with rainfall and topography. The rainfall source term is relatively simpler compared to the topography source term, as it is merely a spatio-temporal function without involving the dependent variables of the equation. Similar to the topography source term, periods of higher rainfall intensity result in larger approximation errors for PINNs. This behavior differs from the error accumulation pattern observed in traditional numerical methods over time. Our training experience indicates that cases incorporating rainfall often exhibit slower convergence during training and require more training iterations. To mitigate the challenges posed by significant rainfall source terms, methods analogous to those used for addressing topography source terms can be employed.

In Section \ref{sec2_3}, we theoretically demonstrate that the variable-conservation form is superior to the variables form because it automatically weights the loss term \(\mathcal{L}_E\) of SWE point by point through matrix \(A\), which is naturally derived from the conservation law. This weighting mechanism enhances the accuracy and stability of the solutions. In the pseudo-2D dam-break case, we further validated this conclusion through empirical results. Although the variable-conservation form requires more computational effort compared to the variables form, the improvements in stability and accuracy justify the additional cost.

Intuitively, we hypothesize that embedding the energy conservation law in PINNs can regularize the approximate solution, leading to better performance. In static cases, the introduction of entropy conservation and entropy stability conditions smooths the approximate solution but does not provide significant advantages. In contrast, in the dynamic case dDB8, the introduction of the entropy stability condition leads to over-smoothing, resulting in training failure. The benefits of incorporating entropy conditions do not outweigh the increased training costs and instability risks they introduce.

As a hyperbolic conservation law, the shallow water equations essentially propagate initial conditions within the characteristic domain. Deviations in the initial conditions will deform and propagate within this domain and generally do not dissipate automatically. Consequently, the initial condition term plays a crucial role in the PINNs loss function. Observations from Table \ref{tab2} indicate that the error in the initial conditions nearly sets the lower bound for the PINNs approximation error; that is, the magnitude of the approximation error at other times is greater than or equal to that of the initial condition error. It should be noted that boundary conditions hold equal importance to initial conditions. However, due to the relatively simple boundary conditions set in this study, their effects have not been fully reflected.

Achieving engineering applications of the PINNs method for solving real-world long-term and large-scale rainfall and flood problems remains a significant challenge. Current difficulties include, but are not limited to, high computational costs, stability issues, complex terrain topologies, long-term simulations, and large-scale problem solving.

\section{Conclusion}
PINNs, as a novel methodology for solving two-dimensional shallow water equations, are attracting increasing attention. This study evaluates the capability of PINNs to solve 2D SWE with topography and rainfall source terms through multiple case studies. The results demonstrate that PINNs can effectively handle the "well-balance" problem for both static water and static water with rainfall source terms, as well as complex dynamic problems involving topography and rainfall. There are many issues regarding the PINNs method that remain to be explored. This study addresses the two fundamental questions. First, compared to the conservative form and variables form, we have theoretically and empirically demonstrated that the variable-conservation form of SWE is a more reliable choice for embedding in PINNs. And from our analysis process, more equation forms can be inspired to adapt to practical problems. Second, our findings indicate that introducing the energy conservation law or entropy condition does not yield significant benefits and may even increase the risk of training failure, contrary to initial expectations. Additionally, this study presents an open-source module developed on the PINNacle platform for solving SWE using PINNs. The module includes over ten case studies and various equation forms, and is equipped with two numerical solvers: the HLL scheme and the entropy stable scheme ES1. These cases could serve as benchmarks for evaluating new PINN architectures.

\section*{Data Availability Statement}
The source code reported in the study is available on \url{https://github.com/tianyongsen/PINN_SWE_open.git}.

\bibliography{references}
\bibliographystyle{plainnat}

\end{document}